\documentstyle[12pt]{article}
\begin{document}

\pagestyle{empty} \noindent
\hspace*{115mm} \normalsize IASSNS-HEP-93/39 \\ \hspace*{115mm} July 1993 \\
\hspace*{115mm} revised September 1993\\
\begin{center} \vspace*{30mm} \LARGE $N=2$ gauged WZW models \linebreak and the
elliptic genus\\
\vspace*{20mm} \large M\aa ns Henningson$^\star$\\
\vspace*{10mm} \normalsize \it School of Natural Sciences \\ Institute for
Advanced Study \\ Olden Lane \\ Princeton, NJ 08540 \\
\vspace*{25mm} \large Abstract \\ \end{center}
Witten recently gave further evidence for the conjectured relationship between
the $A$ series of the $N=2$ minimal models and certain Landau-Ginzburg models
by computing the elliptic genus for the latter. The results agree with those of
the $N=2$ minimal models, as can be calculated from the known characters of the
discrete series representations of the $N=2$ superconformal algebra. The $N=2$
minimal models also have a Lagrangian representation as supersymmetric gauged
WZW models. We calculate the elliptic genera, interpreted as a genus one path
integral with twisted boundary conditions, for such models and recover the
previously known result.

\vspace*{10mm} \noindent
$\star$ Research supported by the Swedish Natural Science Research Council
(NFR).
\newpage \pagestyle{plain} \begin{center}{\Large 1. Introduction}\end{center}

The study of $N=2$ superconformal field theories in two dimensions has played
an increasingly important role in string theory in recent years. The main
motivation has been to construct vacua of string theory which exhibit
space-time supersymmetry, but there are also intriguing connections with
topological field theory. Furthermore, the $N=2$ structure is of great
mathematical interest.

Among representations of the $N=2$ superconformal algebra, those in the
discrete series are of particular importance
\cite{DiVecchia-Petersen}\cite{Boucher-Friedan-Kent}. By assembling left- and
right-moving such representations, we may construct exactly solvable examples
of $N=2$ superconformal field theories, the so called minimal models. Modularly
invariant such theories obey an $ADE$ classification, much as the modular
invariants of affine $SU(2)$ \cite{Cappelli-Itzykson-Zuber}\cite{Gepner-Qiu}.

Several different Lagrangian formulations of $N=2$ superconformal field
theories are known. Important examples include non-linear sigma models with
Calabi-Yau target spaces, supersymmetric gauged WZW models and supersymmetric
Landau-Ginzburg theories. There is a certain overlap between the different
constructions, though.

There are strong reasons to believe that certain supersymmetric Landau-Ginzburg
models flow to the $A$ series of the minimal models in the infrared under the
renormalization group. To test this conjecture one should find quantities which
are effectively computable both for the Landau-Ginzburg models and the minimal
models. An interesting example is the elliptic genus
\cite{Schellekens-Warner}\cite{Witten-87}, which could be viewed as a
restriction of the usual partition function. In contrast to the partition
function, the elliptic genus has an interpretation as an index of one of the
supercharges, and is therefore invariant under a large class of smooth
deformations of the theory. This property was recently used by Witten
\cite{Witten-93} to compute the elliptic genus of the Landau-Ginzburg models by
deforming them to free field theories.

Di Francesco and Yankielowicz \cite{Francesco-Yankielowicz} have shown that the
result of Witten's computation agrees with the elliptic genus of the minimal
models. The proof uses the known characters of the discrete series
representations of the $N=2$ superconformal algebra. However, the $A$ series of
the $N=2$ minimal models also have a Lagrangian representation as
supersymmetric gauged WZW models. To get a better conceptual understanding of
the relationship between the different formulations, it would be valuable to
calculate the elliptic genus of the minimal models in a Lagrangian formalism,
i.~e. as a genus one path integral with twisted boundary conditions. This is
the object of the present paper.

The organization of this paper is as follows: In section two we discuss the
definition and general characteristics of the elliptic genus and review
Witten's calculation \cite{Witten-93} of it for certain Landau-Ginzburg models.
In section three we briefly review the $N=2$ supersymmetric coset models and
their formulation as supersymmetric gauged WZW models. In section four we
calculate the elliptic genera of the $N=2$ minimal models in a path integral
formalism by deforming the models to a weak coupling limit where a one-loop
approximation can be justified.

\newpage \begin{center}{\Large 2. The elliptic genus and Landau-Ginzburg
theory}\end{center}

An important quantity characterizing an $N=2$ superconformal field theory is
the partition function
\begin{equation}
Z(q,\gamma_L,\gamma_R) = {\rm Tr} (-1)^F q^{L_0} {\bar q}^{\bar L_0} \exp
(i\gamma_L J_0 + i\gamma_R {\bar J}_0),
\end{equation}
where $J_0$ and ${\bar J}_0$ are the global $U(1)$ charges of the left- and
right-moving $N=2$ algebra respectively. Although the partition function of for
example the minimal models is known, it is not effectively computable for a
general $N=2$ superconformal field theory.

The situation is much better for the elliptic genus
\cite{Schellekens-Warner}\cite{Witten-87}, which is simply the restriction of
the partition function to $\gamma_R=0$, i.~e. $Z(q,\gamma,0)$. In the case of
for example a sigma model, the elliptic genus is a topological invariant of the
target space. This is related to the fact that it can be interpreted as an
index of the $N=1$ right-moving supercharge, as we will now explain.

The global right-moving $N=1$ supersymmetry algebra is $Q_R^2 = {\bar L}_0$.
The index of $Q_R$, i.~e. the difference between the number of bosonic and
fermionic states of ${\bar L}_0=0$ can be written as
\begin{equation}
{\rm Tr} (-1)^{F_R} {\bar q}^{{\bar L}_0},
\end{equation}
since states of non-zero ${\bar L}_0$ eigenvalue cancel pairwise. Defined in
this way, the index is in general divergent due to the infinite degeneracy of
the left-moving degrees of freedom. This could be remedied by the inclusion of
a convergence factor, which should commute with $Q_R$. A convenient choice for
an $N=2$ theory is
\begin{equation}
(-1)^{F_L} q^{L_0} \exp(i \gamma J_0),
\end{equation}
which gives exactly the elliptic genus. States of non-zero ${\bar L_0}$ still
cancel, so the elliptic genus is holomorphic in $q$, i.~e. ${\bar q}$
independent. For some other general properties, see \cite{Kawai-Yamada-Yang}.

Being an index of the right-moving supercharge, the elliptic genus is invariant
under smooth deformations of the theory which preserve an $N=1$ right-moving
supersymmetry. This property was recently used by Witten \cite{Witten-93} to
calculate the elliptic genus of certain Landau-Ginzburg models. These models
are given by the action
\begin{eqnarray}
I(\phi,\psi) & = & \int d^2 z \left( -\partial_z {\bar \phi} \partial_{\bar z}
\phi + i {\bar \psi}_- \partial_z \psi_- + i {\bar \psi}_+ \partial_{\bar z}
\psi_+ \right.\nonumber\\
&& \left.- ({\bar \phi} \phi )^{k+1} - (k+1) \phi^k \psi_- \psi_+ - (k+1) {\bar
\phi}^k {\bar \psi}_+ {\bar \psi}_- \right),
\end{eqnarray}
where $\phi$ and $\psi$ are complex bosonic and fermionic fields respectively.

The global $U(1)$ symmetry with parameter $\gamma$ which is part of the
left-moving $N=2$ algebra acts as \cite{Witten-93}
\begin{eqnarray}
\phi & \rightarrow & \exp (\frac{i \gamma}{k+2}) \phi \nonumber\\
\psi_+ & \rightarrow & \exp (\frac{i \gamma}{k+2}) \psi_+ \label{LG-susy}\\
\psi_- & \rightarrow & \exp (- \frac{i \gamma (k+1)}{k+2}) \psi_-. \nonumber
\end{eqnarray}

To calculate the elliptic genus we deform the action by turning off the
interactions, which turns the model into a theory of a free complex boson and a
free complex fermion. The left-moving $U(1)$ transformations are still given by
(\ref{LG-susy}). The calculation of the elliptic genus is now easily performed
in an operator formalism. The non-zero modes of $\psi_-$, ${\bar \psi}_-$,
$\psi_+$ and ${\bar \psi}_+$ give rise to the factors
\begin{equation}
\prod_{n=1}^\infty (1-q^n \exp(\!-\!\frac{-i\gamma (k\!+\!1)}{k\!+\!2})) (1-q^n
\exp(\frac{i\gamma (k\!+\!1)}{k\!+\!2})) (1-{\bar q}^n
\exp(\frac{i\gamma}{k\!+\!2})) (1-{\bar q}^n
\exp(\!-\!\frac{-i\gamma}{k\!+\!2})),
\end{equation}
whereas the contribution of the non-zero modes of $\phi$ and ${\bar \phi}$ is
\begin{equation}
\prod_{n=1}^\infty \left[ (1-q^n\exp(\frac{i \gamma}{k\!+\!2})) (1-{\bar
q}^n\exp(\frac{i \gamma}{k\!+\!2})) (1-q^n\exp(\frac{-i \gamma}{k\!+\!2}))
(1-{\bar q}^n\exp(\frac{-i \gamma}{k\!+\!2})) \right]^{-1}.
\end{equation}
The final result, including the zero modes, is \cite{Witten-93}
\begin{eqnarray}
Z(q,\gamma,0) & = &  e^{-\frac{i\gamma k}{2(k+2)}} \frac{1-\exp(\frac{i\gamma
(k+1)}{k+2})}{1-\exp(\frac{i\gamma}{k+2})} \label{ell-genus} \\
&& \times \prod_{n=1}^\infty \frac{\left(1-q^n \exp(\frac{i \gamma
(k+1)}{k+2})\right) \left(1-q^n \exp(-\frac{i\gamma (k+1)}{k+2})\right)}
{\left(1-q^n \exp(\frac{i \gamma }{k+2})\right) \left(1-q^n \exp(-\frac{i\gamma
}{k+2})\right)}. \nonumber
\end{eqnarray}
By using the known character formulas for the $N=2$ discrete series
representations, Di Francesco and Yankielowicz \cite{Francesco-Yankielowicz}
have shown that this result agrees with the elliptic genus of the minimal
models.

\vspace*{15mm} \begin{center}{\Large 3. The Lagrangian formulation of N=2 coset
models}\end{center}

The $N=1$ coset models \cite{Goddard-Kent-Olive} constitute an important class
of superconformal field theories. Such a model is specified by a Lie group $G$
with a subgroup $H$ and a positive integer $k$ called the level. The central
charge of the $N=1$ superconformal algebra is
\begin{equation}
c= \frac{3}{2} ( {\rm dim} G - {\rm dim} H ) - \frac{1}{k+Q_G} ( Q_G \, {\rm
dim} G - Q_H \, {\rm dim} H ), \label{c}
\end{equation}
where $Q_G$ and $Q_H$ are the dual Coxeter numbers of $G$ and $H$ respectively.

Schnitzer \cite{Schnitzer} has given a Lagrangian representation of the coset
models as supersymmetric, gauged WZW models. The fundamental fields are a $G$
valued bosonic field $g$, a gauge field $A_\alpha$ with values in ${\rm Lie} \,
 H$, and left- and right-moving fermionic fields ${\hat \psi}_+$ and ${\hat
\psi}_-$ respectively with values in ${\rm Lie } \,  (G/H)$, i.~e. the
orthogonal complement of ${\rm Lie \,} H$ in ${\rm Lie \,} G$. The action is
\begin{equation}
I(g,A,{\hat \psi}) = I_B(g,A) + I_F(A,{\hat \psi}), \label{action}
\end{equation}
where the bosonic part is given by
\begin{equation}
I_B(g,A) = k I_{WZW}(g) + \frac{k}{2\pi} \int d^2 z \, {\rm Tr}  (A_{\bar z}
g^{-1} \partial_z g - A_z \partial_{\bar z} g g^{-1} - A_{\bar z} A_z + A_{\bar
z} g^{-1} A_z g ) \label{b-action}
\end{equation}
and the fermionic part by
\begin{equation}
I_F(A,{\hat \psi}) = \frac{ik}{4\pi} \int d^2 z \,  {\rm Tr}  ({\hat \psi}_+
D_{\bar z} {\hat \psi}_+ + {\hat \psi}_- D_z {\hat \psi}_-). \label{f0-action}
\end{equation}
Here $I_{WZW}(g)$ is the (level 1) $G$ Wess-Zumino-Witten action
\cite{Witten-84} and the covariant derivative $D_\alpha$ is defined as
$D_\alpha = \partial_\alpha + [ A_\alpha , \;.\; ] \;$.
Infinitesimal gauge transformations with parameter $\Lambda \in {\rm Lie} \,
H$ act as
\begin{eqnarray}
\delta g & = & [\Lambda,g] \nonumber\\
\delta {\hat \psi}_+ & = & [\Lambda,{\hat \psi}_+] \nonumber\\
\delta {\hat \psi}_- & = & [\Lambda,{\hat \psi}_-] \label{gauge}\\
\delta A_\alpha & =  & - D_\alpha \Lambda . \nonumber
\end{eqnarray}
The model is invariant under left- and right-moving supersymmetries, with
parameters $\epsilon_+$ and $\epsilon_-$ respectively, acting as
\begin{eqnarray}
\delta g & = & i \epsilon_- g {\hat \psi}_+ + i \epsilon_+ {\hat \psi}_- g
\nonumber\\
\delta {\hat \psi}_+ & = & \epsilon_- (1-\Pi_H) (g^{-1} D_z g - i {\hat \psi}_+
{\hat \psi}_+) \nonumber\\
\delta {\hat \psi}_- & = & \epsilon_+ (1-\Pi_H) (D_{\bar z} g g^{-1} + i {\hat
\psi}_- {\hat \psi}_-) \label{susy}\\
\delta A_\alpha & = & 0. \nonumber
\end{eqnarray}
Here $\Pi_H$ is the orthogonal projection of ${\rm Lie} \,  G$ on ${\rm Lie} \,
 H$.

Kazama and Suzuki \cite{Kazama-Suzuki} investigated the conditions under which
the $N=1$ coset models actually possess $N=2$ supersymmetry. They found that
this happens exactly when $G/H$ is a K\"ahler space. A more algebraic way to
formulate this condition is as follows: $G/H$ is a K\"ahler space exactly when
${\rm Lie }\,  (G/H)$ can be decomposed as
\begin{equation}
{\rm Lie }\,  (G/H) = T \oplus {\bar T}, \label{Kahler-0}
\end{equation}
where $T$ and ${\bar T}$ are complex conjugate representations of $H$ such that
\begin{equation}
[T,T] \subset T \;\;\;\;\;\;\; [{\bar T},{\bar T}] \subset {\bar T}
\label{Kahler}
\end{equation}
and
\begin{equation}
{\rm Tr}  (uv) = 0 \;\;\;\; {\rm for} \;\; u,v \in T \;\; {\rm or} \;\; u,v \in
{\bar T}. \label{Kahler-2}
\end{equation}
The simplest example of a Kazama-Suzuki model is $G/H=SU(2)/U(1)$. Note that
(\ref{c}) in this case gives the central charges $c=3k/(k+2)$ of the $N=2$
minimal models.

By scrutinizing the action (\ref{action}) with (\ref{b-action}) and
(\ref{f0-action}), we may in fact see that the model indeed has two left-moving
and two right-moving supersymmetries when the Kazama-Suzuki conditions
(\ref{Kahler-0}), (\ref{Kahler}) and (\ref{Kahler-2}) are fulfilled
\cite{Witten-92}\cite{Nakatsu}. Namely, denoting the components of ${\hat
\psi}_\pm$ in $T$ and ${\bar T}$ as $\psi_\pm$ and ${\bar \psi}_\pm$
respectively, we may write the fermionic part (\ref{f0-action}) of the action
as
\begin{equation}
I_F(A,{\hat \psi}) = \frac{ik}{2\pi} \int d^2 z \,  {\rm Tr}  (\psi_+ D_{\bar
z} {\bar \psi}_+ + \psi_- D_z {\bar \psi}_-). \label{f-action}
\end{equation}
We see that there is an $R$-symmetry, i.~e. a symmetry which does not commute
with the supersymmetries (\ref{susy}), such that $\psi_\pm$ and ${\bar
\psi}_\pm$ have charge $+1$ and $-1$ respectively whereas $g$ and $A$ are
uncharged. Furthermore, the condition (\ref{Kahler}) means that the
supersymmetry transformations (\ref{susy}) may be decomposed in two parts that
change the charge by $+1$ and $-1$ respectively. For the right movers this
yields an $N=2$ supersymmetry with parameters $\epsilon_-$ and ${\bar
\epsilon}_-$ acting as
\begin{eqnarray}
\delta g & = & i \epsilon_- g {\bar \psi}_+ + i {\bar \epsilon}_- g \psi_+
\nonumber\\
\delta \psi_+ & = & \epsilon_- \Pi (g^{-1} D_z g - i \psi_+ {\bar \psi}_+ - i
{\bar \psi}_+ \psi_+ ) - i {\bar \epsilon}_- \psi_+ \psi_+ \nonumber\\
\delta {\bar \psi}_+ & = & {\bar \epsilon}_- {\bar \Pi} (g^{-1} D_z g - i
\psi_+ {\bar \psi}_+ - i {\bar \psi}_+ \psi_+ ) - i \epsilon_- {\bar \psi}_+
{\bar \psi}_+ \label{right-susy}\\
\delta \psi_- & = & \delta {\bar \psi}_- = \delta A_\alpha = 0. \nonumber
\end{eqnarray}
Here $\Pi$ and ${\bar \Pi}$ denote the projections on $T$ and ${\bar T}$
respectively. The analogous decomposition for the left movers is
\begin{eqnarray}
\delta g & = & i \epsilon_+ {\bar \psi}_- g + i {\bar \epsilon}_+ \psi_- g
\nonumber\\
\delta \psi_- & = & \epsilon_+ \Pi (D_{\bar z} g g^{-1} +i \psi_- {\bar \psi}_-
+ i {\bar \psi}_- \psi_- ) + i {\bar \epsilon}_+ \psi_- \psi_- \nonumber\\
\delta {\bar \psi}_- & = & {\bar \epsilon}_+ {\bar \Pi} (D_{\bar z} g g^{-1} +
i \psi_- {\bar \psi}_- + i {\bar \psi}_- \psi_- ) + i \epsilon_+ {\bar \psi}_-
{\bar \psi}_- \label{left-susy}\\
\delta \psi_+ & = & \delta {\bar \psi}_+ = \delta A_\alpha  = 0 . \nonumber
\end{eqnarray}

To compute the elliptic genus it is essential to identify the global $U(1)$
symmetry that is part of the left-moving $N=2$ algebra. The condition that this
$U(1)$ transformation commutes with the right-moving supersymmetries
(\ref{right-susy}) means that ${\rm Lie} \, G$ may be graded by the $U(1)$
charge of $g^{-1} D_z g$, and that this equals the $U(1)$ charge of $\psi_+$ or
${\bar \psi}_+$ on each subspace of $T$ or ${\bar T}$ respectively. The
left-moving supersymmetries (\ref{left-susy}) parametrized by $\epsilon_+$ and
${\bar \epsilon}_+$, on the other hand, should have $U(1)$ charge $+1$ and $-1$
respectively. By the same arguments as before we find that ${\rm Lie} \, G$ may
also be graded by the $U(1)$ charge of $D_{\bar z} g g^{-1}$. However, this
$U(1)$ charge differs from that of $\psi_-$ or ${\bar \psi}_-$ by $+1$ or $-1$
respectively.

Henceforth we will only consider the case where $G/H=SU(2)/U(1)$, i.~e. the
$N=2$ minimal models. A $U(1)$ transformation with parameter $\gamma$ of the
left-moving $N=2$ algebra then acts on the fermionic fields as
\begin{eqnarray}
\delta \psi_- & = & i \gamma c_- \psi_- \nonumber\\
\delta \psi_+ & = & i \gamma c_+ \psi_+ \label{f-transf}
\end{eqnarray}
for some real constants $c_-$ and $c_+$. The fields ${\bar \psi}_\pm$ transform
as the complex conjugates of $\psi_\pm$. For the bosonic fields we postulate
the transformation laws
\begin{eqnarray}
\delta g & = & i \gamma ( x_- U g +  x_+ g U ) \nonumber\\
\delta A_\alpha & = & 0 \label{b-transf}
\end{eqnarray}
for some real constants $x_-$ and $x_+$. Here $U \in {\rm Lie} \, SU(2)$ is the
generator of the gauged $U(1)$ factor in $SU(2)$ normalized so that it has
eigenvalues $+1$ and $-1$ when acting on $T$ and ${\bar T}$ respectively in the
adjoint representation of $SU(2)$. The requirement that the $U(1)$
transformation (\ref{f-transf}) and (\ref{b-transf}) of the left-moving $N=2$
algebra commutes with the right-moving supersymmetries (\ref{right-susy})
translates into the condition $-x_+=c_+$. To get the correct charges for the
left-moving supersymmetries (\ref{left-susy}) we must take $x_-=c_--1$.

We get an additional relation between $c_-$, $c_+$, $x_-$ and $x_+$ from the
requirement that the action (\ref{action}) with (\ref{b-action}) and
(\ref{f-action}) be invariant under the $U(1)$ transformation (\ref{f-transf})
and (\ref{b-transf}). The fermionic part (\ref{f-action}) is invariant at the
classical level under the transformations (\ref{f-transf}), but at the quantum
level the symmetry breaks down due to the chiral anomaly. The anomalous
variation of the effective action is
\begin{equation}
\delta I^{eff}_F(A,{\hat \psi}) = i \gamma 2(c_+-c_-) \frac{1}{2\pi} \int d^2 z
\, {\rm Tr}  (U F_{z {\bar z}}), \label{f-anomaly}
\end{equation}
where the $U(1)$ gauge field strength is defined as $F_{z {\bar z}} =
\partial_z A_{\bar z} - \partial_{\bar z} A_z$. We see that the integrated
anomaly is proportional to the first Chern class of the line bundle on which
the $U(1)$ gauge field $A_\alpha$ is a connection. Note that although
(\ref{f-action}) is proportional to the level $k$, the anomaly
(\ref{f-anomaly}), being a one-loop quantum effect, is independent of $k$.

The bosonic part (\ref{b-action}) of the action, on the other hand, is
non-invariant under (\ref{b-transf}) already at the classical level with
\begin{equation}
\delta I_B(g,A) = i \gamma (- x_- - x_+ ) \frac{k}{2\pi} \int d^2 z \, {\rm Tr}
 (U F_{z \bar z}). \label{b-anomaly}
\end{equation}
There are no quantum corrections to this result, though, since we take the
integration measure ${\cal D} g$ in the bosonic path integral to be a product
of a Haar measure, invariant under left and right $G$ multiplication, for each
point on the world sheet. We thus see that the quantum anomaly
(\ref{f-anomaly}) from the fermionic action will cancel against the classical
anomaly (\ref{b-anomaly}) from the bosonic action if we choose $2 ( c_+ - c_- )
= k ( x_- + x_+ )$.

Finally, by means of a gauge transformation (\ref{gauge}) we may choose
$x_-=x_+$. This gauge choice is convenient in that it allows the $U(1)$
transformation (\ref{b-transf}) to have a line of fixed points where $\delta g
= 0$. These fixed points will be important in the next section. Witb a suitable
normalization, our final transformation laws for the global $U(1)$ symmetry of
the left-moving $N=2$ algebra are thus
\begin{eqnarray}
\delta \psi_+ & = & \frac{i \gamma}{k+2} \psi_+ \nonumber\\
\delta \psi_- & = & \frac{i \gamma (k+1)}{k+2} \psi_- \label{U1-transf}\\
\delta g & = & - \frac{i \gamma}{k+2} (Ug + gU) \nonumber\\
\delta A_\alpha & = & 0 \nonumber
\end{eqnarray}
with ${\bar \psi}_\pm$ again transforming as the complex conjugates of
$\psi_\pm$.

\vspace*{15mm} \begin{center}{\Large 4. Path integral calculation of the
elliptic genus} \end{center}

Our object is now to calculate the elliptic genus of a supersymmetric gauged
WZW model based on $G/H=SU(2)/U(1)$, as discussed in the previous section. In a
path integral formulation, this could be interpreted as a genus one
vacuum-to-vacuum amplitude with the boundary conditions twisted by a
left-moving global $U(1)$ transformation (\ref{U1-transf}) along one of the
cycles of the world sheet. If we take the world-sheet to be a torus with
modular parameter $\tau$, the fields should thus obey the following boundary
conditions:
\begin{eqnarray}
g(z+\tau,{\bar z}+{\bar \tau}) & = & \exp (-\frac{i\gamma}{k+2} U ) g(z,{\bar
z}) \exp (-\frac{i\gamma}{k+2} U ) \nonumber\\
\psi_+(z+\tau,{\bar z}+{\bar \tau}) & = & \exp (\frac{i\gamma}{k+2})
\psi_+(z,{\bar z}) \nonumber\\
\psi_-(z+\tau,{\bar z}+{\bar \tau}) & = & \exp (\frac{i\gamma (k+1)}{k+2})
\psi_-(z,{\bar z}) \\
A_\alpha(z+\tau,{\bar z}+{\bar \tau}) & = & A_\alpha(z,{\bar z}) \nonumber
\end{eqnarray}
with corresponding conditions on ${\bar \psi}_\pm(z,{\bar z})$. All fields
should furthermore be invariant under $z \rightarrow z+1$. The elliptic genus
is given as the Euclidean path integral
\begin{equation}
Z(q,\gamma,0) = \int {\cal D} g {\cal D}  {\hat \psi} {\cal D} A \,
e^{-I(g,A,{\hat \psi})}, \label{FPI}
\end{equation}
where the fields obey the above boundary conditions. As usual $q = \exp (2 \pi
i \tau )$.

As previously mentioned, the elliptic genus has an interpretation as an index
of the right-moving supercharge, and is therefore invariant under smooth
deformations of the theory which preserve this supersymmetry. In a path
integral formalism, we describe such a deformation by adding a perturbation of
the form $\lambda I_1(g,A,{\hat \psi}) = \lambda \int d^2 z \, {\cal L}_1$ to
the action in (\ref{FPI}). The perturbation Langrangian ${\cal L}_1$ should
transform into a total derivative under the right-moving supersymmetry. For the
perturbed model, we thus get
\begin{equation}
Z(q,\gamma,0) = \int {\cal D} g {\cal D}  {\hat \psi} {\cal D} A \,
e^{-I(g,A,{\hat \psi}) - \lambda I_1(g,A,{\hat \psi})}. \label{FPI-pert}
\end{equation}

To see that the elliptic genus (\ref{FPI-pert}) is indeed independent of
$\lambda$, we calculate
\begin{equation}
\frac{\partial Z(q,\gamma,0)}{\partial \lambda} = - \int {\cal D} g {\cal D}
{\hat \psi} {\cal D} A \, I_1(g,A,{\hat \psi}) e^{-I(g,A,{\hat \psi}) - \lambda
I_1(g,A,{\hat \psi})}. \label{dZ}
\end{equation}
A formal argument shows that the expression (\ref{dZ}) vanishes. Namely, as
long as the right-moving supersymmetry acts freely, we may introduce a
collective fermionic coordinate $\theta$ for this symmetry. The path integral
measure will thus contain the factor $d \theta$. But the invariance of
$I(g,A,{\hat \psi})$ and $I_1(g,A,{\hat \psi})$ under supersymmetry means that
the integrand is independent of $\theta$, and the integral thus vanishes by the
rules of Grassmann integration. This argument breaks down, though, if the
supersymmetry has a fixed point \cite{Witten-92}, as turns out to be the case
in our application. The path integral (\ref{FPI-pert}) will then receive
contributions from field configurations in a neighbourhood of the fixed point,
unless the integrand is zero there. To ensure that the elliptic genus is
independent of $\lambda$, we should therefore require $I_1(g,A,\psi)$ to vanish
at the fixed point of the right-moving supersymmetry.

Path integrals such as (\ref{FPI}) are usually hard to calculate, except at
weak coupling where perturbation theory may be used. Our model is certainly
strongly coupled, though, so a direct evaluation seems quite difficult.
However, by perturbing the model in a suitable way we may take it into the weak
coupling regime, and, as argued in the previous paragraph, the elliptic genus
is invariant under supersymmetric such perturbations. To construct such a
supersymmetric perturbation of the Lagrangian, we note that the global
right-moving supersymmetry algebra (\ref{right-susy}) contains two supercharges
$Q_+$ and ${\bar Q}_+$ obeying $\{ Q_+ , Q_+ \} = \{ {\bar Q}_+ , {\bar Q}_+ \}
= 0$ and $\{ Q_+ , {\bar Q}_+ \} = 2 D_z$. If we take $V$ such that $\{ {\bar
Q}_+ , V \} = 0$, then the perturbation ${\cal L}_1 = \{ Q_+ , V \}$ will be
invariant under $Q_+$ and transform into a total derivative under ${\bar Q}_+$,
i.~e. $[ Q_+ , {\cal L}_1 ] = 0$ and $[ {\bar Q}_+ , {\cal L}_1 ] = 2 D_z V$.

Since we want ${\cal L}_1$ to be Grassmann even, uncharged under the
right-moving $U(1)$ symmetry and of scaling dimension $(1,1)$, we see from
(\ref{right-susy}) that $V$ should be Grassmann odd, of right $U(1)$ charge
$-1$ and of scaling dimension $(1,1/2)$. In our case there is indeed a sensible
such term, namely $V = {\rm Tr} ( g^{-1} D_{\bar z} g \psi_+ )$. With this $V$
we get
\begin{equation}
{\cal L}_1 = {\rm Tr} \left( g^{-1} D_{\bar z} g \Pi (g^{-1} D_z g) + i D_{\bar
z} {\bar \psi}_+ \psi_+ + i g^{-1} D_{\bar z} g (\psi_+ {\bar \psi}_+ + {\bar
\psi}_+ \psi_+ ) \right).
\end{equation}
Note that at a fixed point of the right-moving supersymmetry
(\ref{right-susy}), $g$ must be a constant and $\psi_+$ and ${\bar \psi}_+$
must vanish. This means that ${\cal L}_1$ vanishes, so the elliptic genus is
indeed invariant under this perturbation according to our previous arguments.

To see that the addition of this term to the action takes us to a weakly
coupled theory, it is convenient to change variables from $g \in SU(2)$ to
$\phi \in {\rm Lie \,} SU(2)$ defined through
\begin{equation}
g = g_0 \exp(i\phi),
\end{equation}
where $g_0 \in SU(2)$ is a constant which is invariant under the left-moving
$U(1)$ transformations (\ref{U1-transf}), i.~e.
\begin{equation}
U g_0 + g_0 U = 0.
\end{equation}
Expanding ${\cal L}_1$ we get
\begin{equation}
{\cal L}_1 ={\rm Tr} \left( - D_{\bar z} \phi \Pi (D_z \phi ) + i D_{\bar z}
{\bar \psi}_+ \psi_+ - D_{\bar z} \phi (\psi_+ {\bar \psi}_+ + {\bar \psi}_+
\psi_+ ) \right)
\end{equation}
plus higher order interaction terms.

Under a left-moving $U(1)$ transformation (\ref{U1-transf}) $\phi$ transforms
as
\begin{equation}
\delta \phi = - \frac{i \gamma}{k+2}[U,\phi] + {\cal O}(\phi^2).
\end{equation}
The components of $\phi$ which lie in $T$ and $\bar T$ thus have the same
$U(1)$ charges as ${\bar \psi}_+$ and $\psi_+$ respectively. The remaining
component of $\phi$ in ${\rm Lie \, } H$, i.~e. along the gauged $U(1)$
subgroup of $SU(2)$, has zero charge under the left-moving $U(1)$
transformation, but is in fact pure gauge. To see this, we note that under a
gauge transformation (\ref{gauge}) $\phi$ transforms as
\begin{equation}
\delta \phi = - i (\Lambda - g_0 \Lambda g_0^{-1} ) + {\cal O}(\phi).
\end{equation}
Furthermore, for $\Lambda \in {\rm Lie \, } H$,
\begin{equation}
-\frac{i \gamma}{k+2} [ U , \Lambda - g_0 \Lambda g_0^{-1} ] = 0,
\end{equation}
so the component of $\phi$ which is pure gauge indeed has zero $U(1)$ charge,
and must therefore belong to ${\rm Lie \, } H$.

The K\"ahler condition (\ref{Kahler}) means that $\psi_+ {\bar \psi}_+ + {\bar
\psi}_+ \psi_+  \in {\rm Lie \, } H$. The ${\rm Tr } ( D_{\bar z} \phi (\psi_+
{\bar \psi}_+ + {\bar \psi}_+ \psi_+ ))$ term in the action may therefore be
gauged away, since it only contains components of $\phi$ that are pure gauge.
To lowest non-vanishing order, the Lagrangian ${\cal L}_1$ thus describes two
free bosons and two free fermions.

We may now calculate the elliptic genus in a path integral formalism as a
perturbative expansion in the number of loops. This is tantamount to expanding
the action of the model around its minima, i.~e. around field configurations
which solve the classical equations of motion. In the weak coupling limit,
i.~e. as $\lambda \rightarrow \infty$, the integrand in (\ref{FPI-pert}) is
sharply peaked at the classical field configurations and we may trust the
one-loop or Gaussian approximation, where we expand the action to second order
in the fields around the minima. In this approximation the calculation amounts
to a free-field computation, much as in section two. We see that the essential
feature is the charges of the various fields under the left-moving $U(1)$
transformation, and since these agree with those of the Landau-Ginzburg model
(confer (\ref{LG-susy})) in section two we recover the result (\ref{ell-genus})
of \cite{Witten-93}\cite{Francesco-Yankielowicz} for the elliptic genus of the
$N=2$ minimal models

{}.

\vspace*{3mm}
I would like to thank E. Witten for suggesting this problem and for many
helpful conversations.

\end{document}